%% file: charm2015_SnehaMalde.tex
\documentclass[12pt]{article}
\usepackage{graphicx}

\newcommand\pubnumber{WSU--HEP--XXYY}
\newcommand\pubdate{\today}

\def\oxford{Department of Physics\\
University of Oxford, Oxford, OX1 3RH, UK}
\def\support{\footnote{Work supported by the UK-India Education and Research Initiative, UK Science and Technology Facilities Council and L'Oreal-Unesco For Women In Science.}}

\def\fplus{$F_+^{4\pi}$}
\def\CP{$CP$}
\def\fpion{$D \to \pi^+\pi^-\pi^+\pi^-$}
\def\Title#1{\begin{center} {\Large #1 } \end{center}}
\def\Author#1{\begin{center}{ \sc #1} \end{center}}
\def\Address#1{\begin{center}{ \it #1} \end{center}}

\newcommand\pubblock{\rightline{\begin{tabular}{l} \pubnumber\\
         \pubdate  \end{tabular}}}
\newenvironment{Abstract}{\begin{quotation}  }{\end{quotation}}
\newenvironment{Presented}{\begin{quotation} \begin{center} 
             PRESENTED AT\end{center}\bigskip 
      \begin{center}\begin{large}}{\end{large}\end{center} \end{quotation}}
\def\Acknowledgements{\bigskip  \bigskip \begin{center} \begin{large}
             \bf ACKNOWLEDGEMENTS \end{large}\end{center}}

\input econfmacros.tex

\begin{document}
\begin{titlepage}
\pubblock

\vfill
\Title{First measurement of $F_+^{4\pi}$ in \fpion decays and a new method for measuring \CP-Violation in charm decays}
\vfill
\Author{Sneha Malde\support}
\Address{\oxford}
\vfill
\begin{Abstract}
A first measurement of \fplus, the \CP-even content of the decay \fpion using decays of $\psi(3770)$ to quantum correlated $D\bar{D}$ decays collected by the CLEO-c experiment is presented. The measured value is \fplus=$0.737 \pm 0.028$. This relatively high value makes the decay mode particularly suitable for both measurements of the CKM angle $\gamma$ and charm \CP-violation in a model independent way. This novel approach to studying indirect \CP-violation in charm decays is based on the time-dependent inclusive analysis of multi body self-conjugate states. These final states can be used to determine the indirect \CP-violating observable $A_{\Gamma}$ and the mixing observable $y_{CP}$ provided that $F_+$ is known. This approach can yield significantly improved sensitivity compared with the conventional method that relies on decays to \CP~eigenstates.  
\end{Abstract}
\vfill
\begin{Presented}
The 7th International Workshop on Charm Physics (CHARM 2015)\\
Detroit, MI, 18-22 May, 2015
\end{Presented}
\vfill
\end{titlepage}
\def\thefootnote{\fnsymbol{footnote}}
\setcounter{footnote}{0}
%

\section{Introduction}

The search for physics beyond the standard model (SM) continues to intensify with the continued lack of evidence from direct searches. One avenue is precision measurements of \CP-violation. The CKM angle $\gamma$ is the only angle that is accessible at tree level and hence direct measurements form a SM benchmark. Discrepancy between direct and indirect measurements, that can contain loop processes, would be a sign of new physics(NP). \CP-violation in the charm sector is also of great interest. The SM predicts very small levels of \CP-violation in the charm sector, and hence any observation with current data sets could imply NP~\cite{CharmCPV}. In both the beauty and charm sector better precision is required to observed deviations from expectation. One way of increasing precision is to increase the range of decay modes that can be used to make measurements. The program of analysis of CLEO-c data to provide information on the \CP~content or strong phases of $D$ decays provides the input necessary to use the decay modes in \CP-violation measurements.\\

Recent measurements of the \CP-even content of $D \to \pi\pi\pi^0$ and $D \to K K \pi^0$~\cite{hhpi0} have discussed how knowledge of these parameters can benefit measurements of the CKM angle $\gamma$. As discussed in Ref.~\cite{fpluscpv}, and in these proceedings, the knowledge of the \CP~content of these self conjugate states can also be used to improve the precision of the charm \CP-violating observables $A_{\Gamma}$ and $y_{CP}$. There are further self conjugate $D$ meson decays that could display a high \CP-even or \CP-odd content. These proceedings describe one such new measurement of the \CP-even content of \fpion.

\section{Measurement of \fplus in \fpion decays}

The \CP-content of self-conjugate $D$ meson decays can be determined from analysis of quantum correlated $D^0\bar{D^0}$ pairs originating from the $\psi(3770)$ resonance, collected by the CLEO-c detector. 
The analysis exploits events where one $D$ meson is reconstructed in the signal mode ($D \to 4\pi$) and the other $D$ meson (the tag) is reconstructed in a decay to a \CP~eigenstate, or other tags where the \CP~content is known, either over the full phasespace ($D \to \pi\pi\pi^0$) or only in a particular region ($D \to K^0_{S,L} \pi\pi$). The full details of the analysis are now published, and full analysis details are available in Ref.~\cite{4pi}.

The data set analysed consists of $e^+e^-$ collisions produced by the Cornell Electron Storage Ring (CESR) at $\sqrt{s} = 3.77$ GeV corresponding to an integrated luminosity of 818 pb$^{-1}$ and collected with the CLEO-c detector~\cite{cleodet}. The reconstruction efficiencies and backgrounds are determined where necessary from simulated samples. 
The $D$ meson final states considered in the analysis are listed in Table~\ref{tab:modes}. The detector performance results in selected samples of high purity ~\cite{4pi}. 
\begin{table}[th]
\begin{center}
\label{tab:modes}
\begin{tabular}{c|c}
Type & Final states \\ \hline
Mixed \CP~& $\pi^+\pi^- \pi^+\pi^-$, $\pi^+\pi^-\pi^0$, $K^0_{S, L}\pi^+\pi^-$ \\
\CP-even & $K^{+}K^{-}$, $\pi^{+}\pi^{-}$, $K^{0}_{S}\pi^{0}\pi^{0}$, $K^{\
0}_{L}\pi^{0}$, $K^{0}_{L}\omega$ \\
\CP-odd  & $K^{0}_{S}\pi^{0}$, $K^{0}_{S}\omega$, $K^{0}_{S}\eta$, $K^{0}_{S}\eta^{\prime}$ \\
\hline
\end{tabular}
\caption{$D$-meson final states reconstructed in this analysis.} \vspace*{0.1cm\
}
\end{center}
\end{table}

\begin{figure}[htb]
\centering
\includegraphics[height=2.5in]{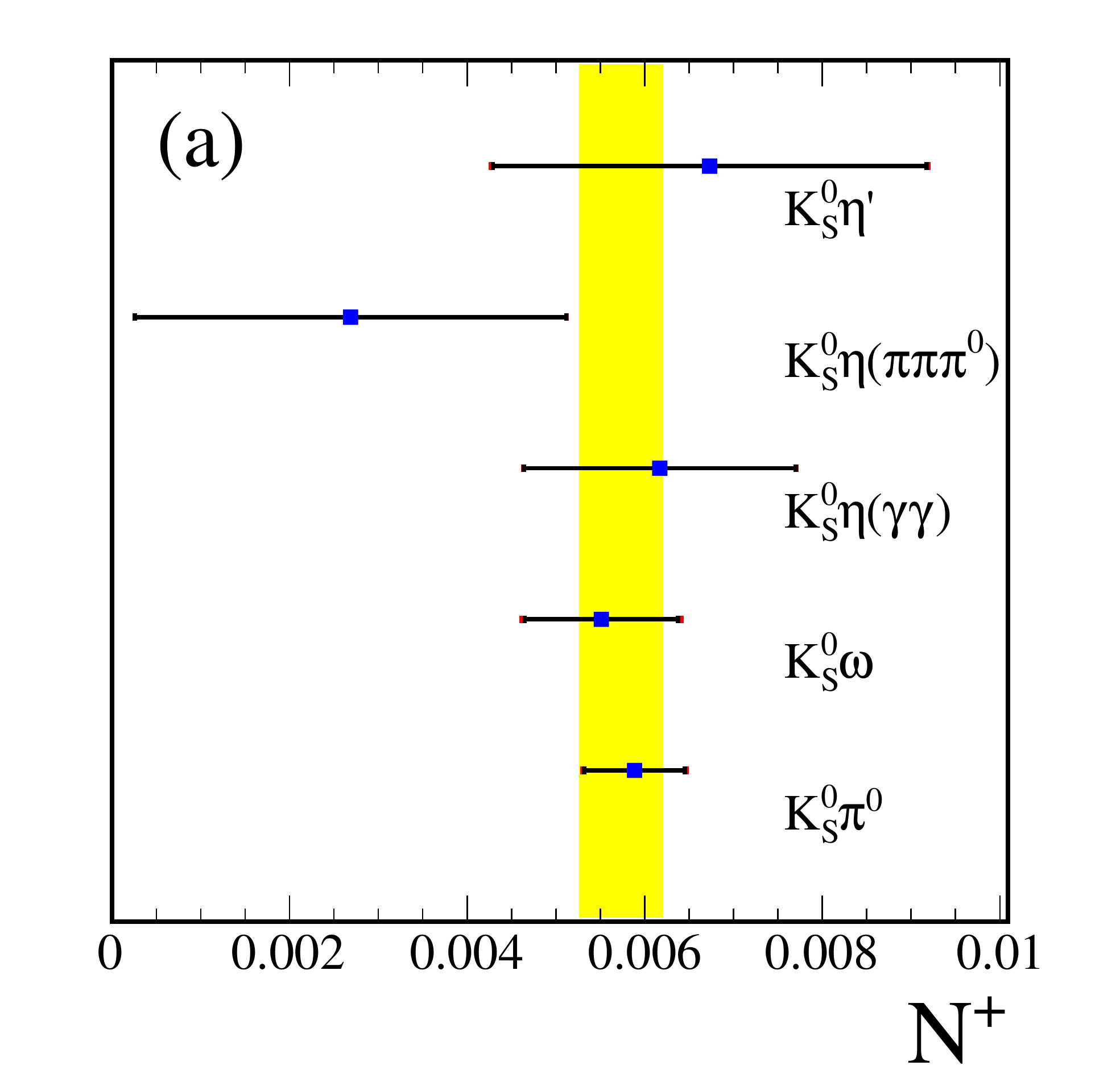}
\includegraphics[height=2.5in]{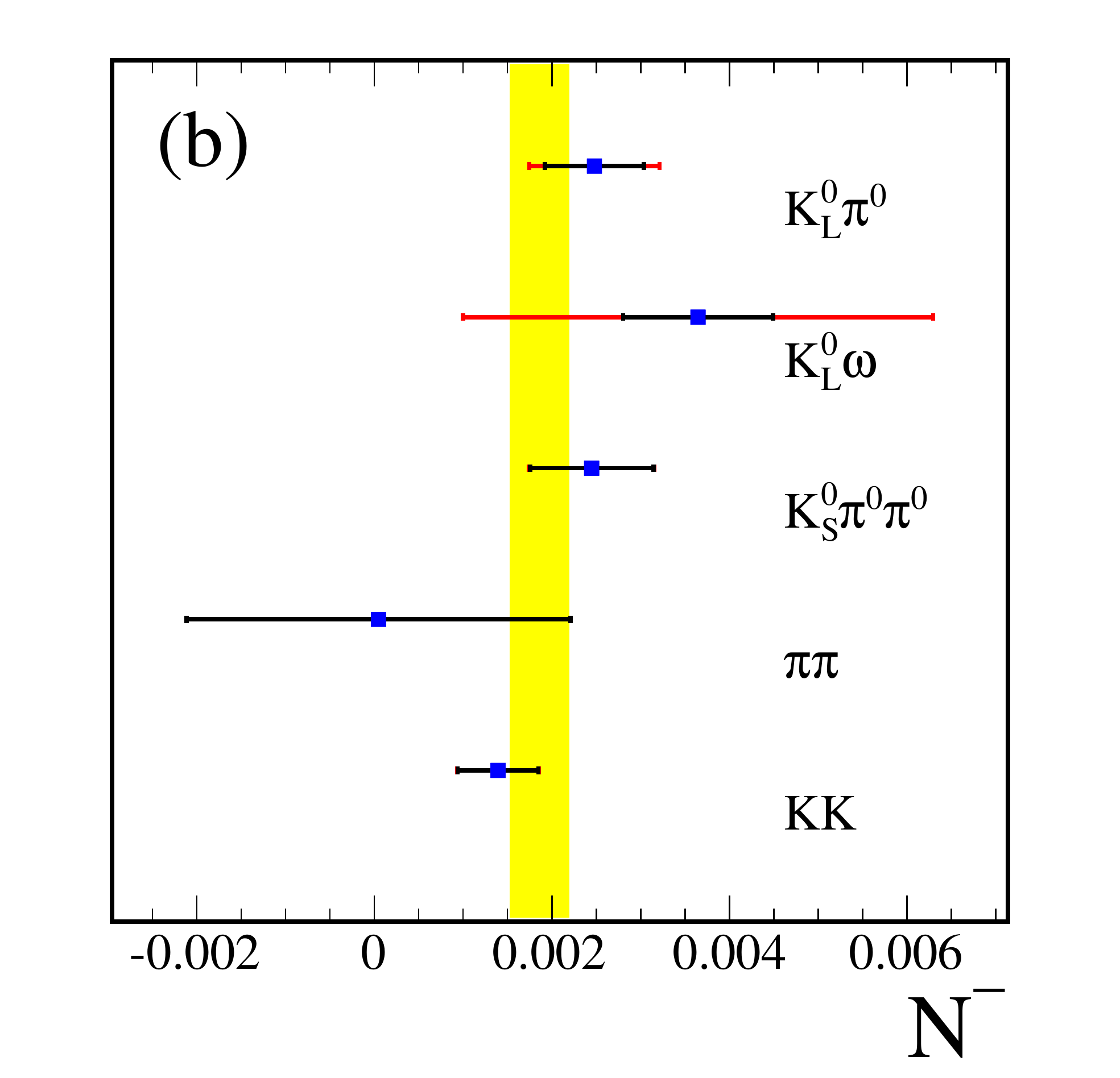}
\caption{$D \to \pi^+\pi^-\pi^+\pi^-$ results for (a) $N^+$ and (b) $N^-$. In each plot the vertical yellow band indicates the value obtained from the combination of all tags. The black portion of the uncertainty represents the statistical uncertainty only while the red represents the total. }
\label{fig:nplus}
\end{figure}

In this analysis the double tag yield $(M)$ refers to the signal yield where one $D$ meson decays to $4\pi$ and the final state of the other $D$ meson in the event is determined to be one of the final states in Table~\ref{tab:modes}. The single tag yield refers to the yield of any of the final states in Table~\ref{tab:modes} where no criteria are placed on the decay of the other $D$ meson. It can be shown~\cite{4pi} that \fplus is given by
\begin{equation}
F_+^{4\pi} = \frac{N^+}{N^+ + N^-}
\end{equation}
where $N^+$ is the ratio of the double tag yield with a \CP-odd eigenstate tag to the single tag yield of the \CP-odd eigenstate. The observable $N^-$ is analogous for the \CP-even eigenstate tags. The ratio is taken to remove dependence on \CP-tag branching fractions and reconstruction efficiencies. In the case of the \CP-tags including $K_L^0$ the effective single tag yield is calculated from branching fractions, reconstruction efficiencies and the number of $D\bar{D}$ pairs in the dataset. Due to this, the systematic uncertainties for these tags are larger. The values of $N^+$ ($N^-$) are consistent between different \CP~tags, as shown in Fig.~\ref{fig:nplus}. The mean value $<N^+>$ = $(5.54 \pm 0.46) \times 10^{-3}$ is significantly larger than $<N^->$ = $(1.80 \pm 0.32) \times 10^{-3}$ indicating that the $\pi^+\pi^-\pi^+\pi^-$ final state is predominantly \CP-even. Using these mean values, and accounting for correlations between systematic uncertainties gives \fplus = 0.754 $\pm$ 0.031 $\pm$ 0.021, where the first uncertainty is statistical and the second systematic.

It is also possible to use tags of mixed \CP~to determine \fplus. The tag $\pi\pi\pi^0$ is used in a similar way to the \CP~tags by constructing the ratio of double tags to single tags $(N^{\pi\pi\pi^0})$ to eliminate uncertainties due to reconstruction and branching fractions. Here, knowledge of $F_+^{\pi\pi\pi^0}$ is required. 
\begin{equation}
F_+^{4\pi}= \frac{N^+F^{\pi\pi\pi^0}_+}{N^{\pi\pi\pi^0} - N^+ + 2N^+F^{\pi\pi\pi^0}_+}.
\end{equation}
Using the value of $F_+^{\pi\pi\pi^0}$ from ~\cite{hhpi0},  $F_+^{4\pi} = 0.0765 \pm 0.050$.

\begin{figure}[htb]
\centering
\includegraphics[height=2.5in]{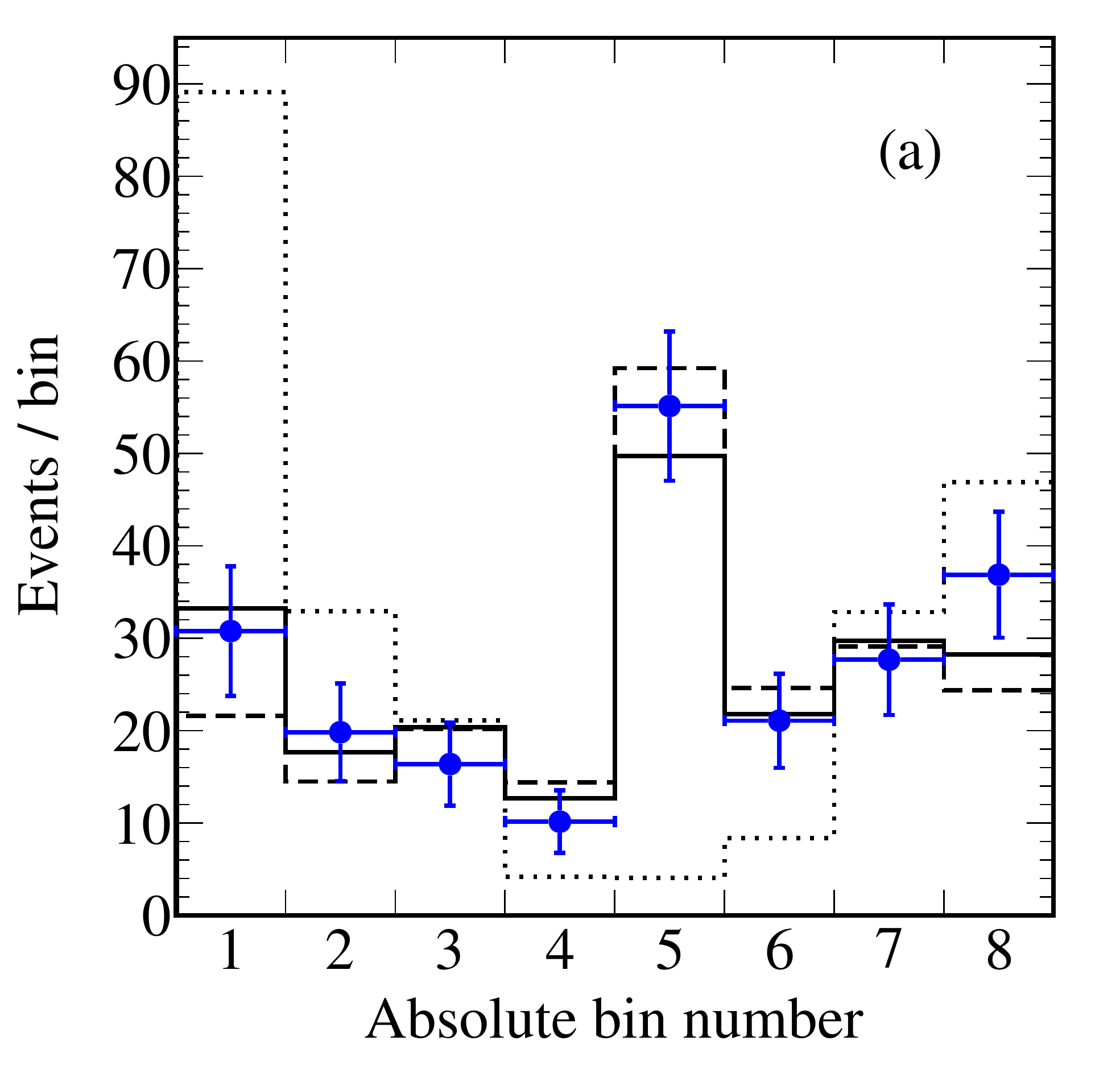}
\includegraphics[height=2.5in]{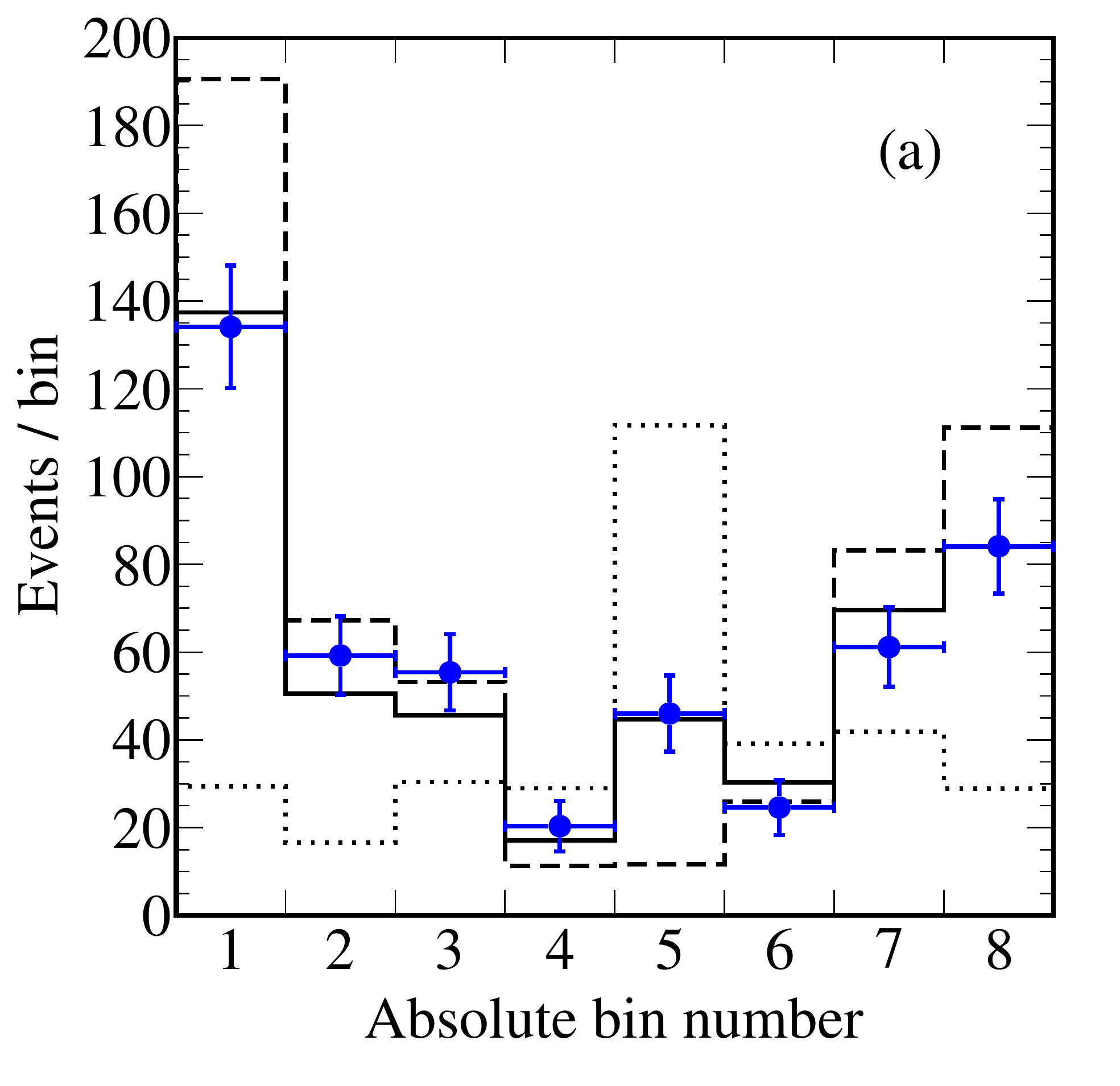}
\caption{Data (points) and fit results (solid line) in absolute bin numbers for $K_S^0\pi\pi$ (left) and $K_L^0 \pi\pi$ tags. Also shown in each case is the expectation if \fplus =0 (dotted line) or \fplus =1 (dashed line). }
\label{fig:mixedcp}
\end{figure}

 The final tag category that is considered is $K_{S,L}^0\pi\pi$, where the double tagged events are split into bins of the Dalitz plot of the $K_{S,L}^0\pi\pi$ decay. The bins used are the 'equal binning' scheme described in ~\cite{Kspipi} where the measurements of $c_i$, ($c_i'$) the cosine of the strong phase difference weighted by the decay rate of $D \to K_S^0\pi\pi (K_L^0\pi\pi)$~\cite{Kspipi} are reported. Following the derivation in Ref.~\cite{4pi} the expected population of bin $|i|$ for double tagged events with $K_S^0\pi\pi$ is given by
\begin{equation}
M_{|i|} = h [K_i + K_{-i} - (2F_+^{4\pi} -1)2c_i\sqrt{K_iK_{-i}}],
\end{equation}
where h is a normalisation factor and $K_i$ is the flavour-tagged fraction, which is the proportion of $D^0 \to K_S^0\pi\pi$ decays that fall into bin $i$. A similar expression exists for the $K_L\pi\pi$ decays~\cite{4pi}. A fit is performed to the background subtracted, efficiency corrected, double tagged yields to determine the best-fit value of \fplus. The values of $c_i$ and $c_i'$ are measured by the CLEO collaboration ~\cite{Kspipi}, the values of $K_i$ are taken from analysis of various $B$-factory models presented in Ref.~\cite{k3pi} and the values of the $K_i'$ are measurements performed with CLEO-c data~\cite{something}. The fit parameters are the normalisation and \fplus, with other parameters Gaussian constrained to their measured values. From these tags the result $F_+^{4\pi} = 0.737 \pm 0.049 \pm 0.024$ is obtained where the first uncertainty is statistical and the second systematic. The results are plotted in Fig.~\ref{fig:mixedcp}, where it is clear that the fit quality is good and that the data are inconsistent with the expectation of $F_+=0$ or $F_+ =1$.

The results from all three tag types are combined taking into account all correlations between sources of uncertainty. The result is 
\begin{equation}
F_+^{4\pi} = 0.737 \pm 0.028.
\end{equation}

\section{\CP-violation in charm decays}

In the SM, indirect \CP-violation in charm decays is expected to be well below the level of precision that can be currently achieved~\cite{CharmCPV}. The current measurements are consistent with no \CP-violation. However many models of NP predict enhancements~\cite{CharmCPV} and hence expanding the repertoire of possible measurements is crucial to exploit all available data. 

The leading observable, $A_{\Gamma}$, is measured from a difference in lifetimes of the decays of the $D^0$ and $\bar{D^0}$ to a \CP~eigenstate. The reliance on \CP~eigenstates effectively reduces the available decay modes to $KK$, $\pi\pi$ and the \CP-odd component of $K_SKK$ since other \CP~eigenstates are experimentally difficult due to presence of $K_L$ or other particles with low reconstruction efficiencies. Therefore, if the multibody self-conjugate decays that have either low or high values of $F_+$ can also be used, this is of benefit to indirect \CP~violation studies in charm decays.

 Defining the equivalent observable to $A_{\Gamma}$ for these non \CP-eigenstates as 
\begin{eqnarray}
 A_\Gamma^{eff} \equiv  \frac{ \hat{\Gamma} - \hat{\bar{\Gamma}} }{\hat{\Gamma} + \hat{\bar{\Gamma}}} \nonumber \\
\end{eqnarray}
then following the derivation in \cite{fpluscpv}
 \begin{eqnarray}
A_\Gamma^{eff}  &\approx& \frac{1}{2}  (2F_+-1)y \cos\phi_{CP} \left(r_{CP} - \frac{1}{r_{CP}}  \right) -  \nonumber  \\
&& \frac{1}{2}(2F_+-1)x \sin\phi_{CP} \left(r_{CP}+ \frac{1}{r_{CP}}\right). \label{eq:aeff}
\end{eqnarray}
where $x,\ y$ are the usual mixing parameters and $r_{CP}$ and $\phi_{CP}$ define the magnitude and phase of the indirect \CP-violation. Therefore the relation between $A_\Gamma^{eff}$ and  $A_{\Gamma}$ is given by
\begin{equation}
 A_\Gamma^{eff} =\frac{A_{\Gamma}}{2F_+ -1}.
\end{equation}
If the value of $F_+$ is 0 or 1 then the expression reduces to $A_{\Gamma}^{eff}\equiv A_{\Gamma}$ as expected with \CP~eigenstates. On the other hand if $F_+ = 0.5$ there is no sensitivity to the parameters of interest, and hence this method is not useful for self conjugate modes such as $D \to K_S \pi\pi$ where the value of $F_+$ is expected to be close to 0.5.

The $\pi\pi\pi^0$ decay mode is therefore of considerable interest due to its large branching fraction~\cite{PDG} and large value of $F_+$~\cite{hhpi0}, particularly at $e^+e^-$ experiments such as Belle-II where the $\pi^0$ reconstruction efficiency should be good. The $\pi\pi\pi\pi$ decay mode is also of interest as it has a relatively high branching fraction~\cite{PDG} and the fully charged final state should have high reconstruction efficiency. The sensitivities of these channels are compared to the established $D\to KK$ and $D\to \pi\pi$ decay modes in Table 1, assuming the same reconstruction efficiency for all. A further mode that could have high potential is the $D^0 \to K_S\pi\pi\pi^0$ due to the very high branching fraction of over 5$\%$~\cite{PDG} and the presence of \CP-odd eigenstates $K_S\eta$ and $K_s \omega$ as sub modes, however measurement of $F_+$ in this decay mode is required before it could be used in this way.
Similar arguments can also be applied to use these self-conjugate modes to measure $y_{CP}$ where there is a similar dilution factor of ($2F_+-1$)~\cite{fpluscpv}. While the discussion here has been limited to indirect \CP~violation in $D$ decays, extensions to include direct \CP-violation are also possible~\cite{fpluscpv}.
\begin{table}[t]
\begin{center}
\begin{tabular}{l|cccc}  

& \hspace{0.1cm}$K^+K^-$\hspace{0.1cm} & \hspace{0.1cm}$\pi^+\pi^-$\hspace{0.1cm} & \hspace{0.1cm}$\pi^+\pi^-\pi^0$\hspace{0.1cm} & \hspace{0.1cm}$\pi^+\pi^-\pi^+\pi^-$\hspace{0.1cm} \\ \hline
$BF$ $[\times 10^{-2}]$ & $0.396$ & $0.1402$ &  $1.43$  &  $0.742$ \\
$F_+$ & 1 & 1 &  0.973  & 0.737 \\
Uncertainty & 1 & 1.68 & 0.56 & 1.54 \\ \hline
\end{tabular}
\caption{Relative uncertainties on $A_{\Gamma}$ and $y_{CP}$ for the multibody decay modes compared to the \CP~eigenstates, assuming the measured central values of the branching fractions (BF) and the \CP-even fractions $F_+$. The uncertainties are all normalised to that of $D \to K^+ K^-$.}
\end{center}
\end{table}

\section{Conclusions}

The program of measurements using decays of the $\psi(3770)$ to quantum-correlated $D$ meson pairs continues to open up new channels that can be used to measure the CKM angle $\gamma$ in a model-independent way. The measurement of the \CP-even content of the $D \to \pi^+\pi^-\pi^+\pi^-$ decay is one such example. Furthermore the use of these measurements is expanded by their potential to make measurements of \CP-violation in charm mesons. This presents exciting opportunities over the next decade where extremely large samples of beauty and charm decays are expected to be recorded at LHCb and Belle-II. For the best precision, and hence best chance at observing the effects of NP, improved precision of these charm parameters using the quantum-correlated dataset at BES-III will be essential.

\Acknowledgements

I would like to thank the UK-India Education and Research Initiative and the L'Oreal-Unesco For Women In Science fellowship programme for making attendance at this conference possible.


\end{document}

%% file: econfmacros.tex
\def\beq{\begin{equation}}
\def\eeq#1{\label{#1}\end{equation}}
\def\eeqn{\end{equation}}

\def\beqa{\begin{eqnarray}}
\def\eeqa#1{\label{#1}\end{eqnarray}}
\def\eeqan{\end{eqnarray}}

\let\bar=\overbar

\def\Dslash{\not{\hbox{\kern-4pt $D$}}}
\def\dslash{\not{\hbox{\kern-2pt $\del$}}}

\def\msb{{\bar{\ssstyle M \kern -1pt S}}}

